%% file: main.tex
\documentclass{article}

\usepackage{arxiv}

\usepackage[utf8]{inputenc} 
\usepackage[T1]{fontenc}    
\usepackage{hyperref}       
\usepackage{url}            
\usepackage{booktabs}       
\usepackage{amsfonts}       
\usepackage{nicefrac}       
\usepackage{microtype}      
\usepackage{lipsum}		
\usepackage{graphicx}
\usepackage{doi}
\usepackage{array}
\usepackage{subcaption}
\usepackage{tikz}
\usepackage{float}

\newcommand*\circled[1]{\tikz[baseline=(char.base)]{
            \node[shape=circle,fill,inner sep=1pt] (char) {\textcolor{white}{#1}};}}

\usepackage{background}

\backgroundsetup{
  scale=1,
  color=gray,
  opacity=0.4,
  angle=0,
  position=current page.south, 
  vshift=15pt,   
  contents={\scriptsize \textsf{© 2025 Pragmatic Semiconductor Ltd. This work is made available for non-commercial, academic use only. All rights reserved.}}  
}

\title{Flexing RISC-V Instruction Subset Processors to Extreme Edge}

\author{ \href{https://orcid.org/0009-0009-3808-0345}
{\includegraphics[scale=0.06]{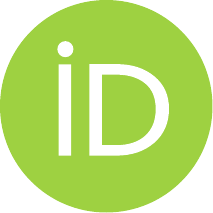}\hspace{1mm}Alireza Raisiardali} \thanks{Also affiliated with Katholieke Universiteit Leuven} \\
    Pragmatic Semiconductor Ltd \\
    400 Cambridge Science Park \\
    Milton Road, Cambridge, CB4 0WH, UK \\
	\texttt{araisiardali@pragmaticsemi.com} \\
    \And
	\href{https://orcid.org/0000-0001-9568-2638}{\includegraphics[scale=0.06]{orcid.pdf}\hspace{1mm}Konstantinos Iordanou} \\
	Pragmatic Semiconductor Ltd \\
    400 Cambridge Science Park \\
    Milton Road, Cambridge, CB4 0WH, UK \\
	\texttt{kiordanou@pragmaticsemi.com} \\
    \And
	{\hspace{1mm}Jedrzej Kufel} \\
	Pragmatic Semiconductor Ltd \\
    400 Cambridge Science Park \\
    Milton Road, Cambridge, CB4 0WH, UK \\
	\texttt{jkufel@pragmaticsemi.com} \\
    \And
    {\hspace{1mm}Kowshik Gudimetla} \\
	Pragmatic Semiconductor Ltd \\
    400 Cambridge Science Park \\
    Milton Road, Cambridge, CB4 0WH, UK \\
	\texttt{gkowshik@pragmaticsemi.com} \\
    \And
    \href{https://orcid.org/0000-0002-5230-495X}
    {\includegraphics[scale=0.06]{orcid.pdf}\hspace{1mm}Kris Myny} \\
	Katholieke Universiteit Leuven \\
    Computer Security and Industrial Cryptography (COSIC) \\
    Diepenbeek Campus, Wetenschapspark 27 - box 15152 \\
    3590 Diepenbeek, Leuven, Belgium \\
	\texttt{kris.myny@kuleuven.be} \\
    \And
    \href{https://orcid.org/0000-0001-8285-1551}{\includegraphics[scale=0.06]{orcid.pdf}\hspace{1mm}Emre Ozer} \\
	Pragmatic Semiconductor Ltd \\
    400 Cambridge Science Park \\
    Milton Road, Cambridge, CB4 0WH, UK \\
	\texttt{eozer@pragmaticsemi.com} \\
}

\renewcommand{\headeright}{}
\renewcommand{\undertitle}{}
\renewcommand{\shorttitle}{Flexing RISC-V Instruction Subset Processors to Extreme Edge}

\hypersetup{
pdftitle={Flexing RISC-V Instruction Subset Processors to Extreme Edge},
pdfauthor={Alireza Raisiardali, Konstantinos Iordanou, Jedrzej Kufel, Kowshik Gudimetla, Kris Myny, Emre Ozer},
pdfkeywords={RISC-V, Flexible electronics, FlexIC,Edge Computing},
}

\begin{document}

\noindent\textbf{Note:} This paper has been accepted for publication at the 58th IEEE/ACM International Symposium on Microarchitecture (MICRO '25) and is available at: \url{https://doi.org/10.1145/3725843.3756036}

\maketitle

\begin{abstract}
This paper presents an automated approach for designing processors that support a subset of the RISC-V instruction set architecture (ISA) for a new class of applications at Extreme Edge. The electronics used in extreme edge applications must be area and power-efficient, but also provide additional qualities, such as low cost, conformability, comfort and sustainability. Flexible electronics, rather than silicon-based electronics, will be able to meet the above qualities. For this purpose, we propose a methodology for generating RISC-V instruction subset processors (RISSPs) tailored to these applications and implementing them as flexible integrated circuits (FlexICs). The methodology makes verification an integral part of the processor design by treating each instruction in the ISA as a discrete, fully functional, pre-verified hardware block. It automatically builds a custom processor by stitching together the instruction hardware blocks required by an application or a set of applications in a specific domain. We generate RISSPs using the proposed methodology for three extreme edge applications, and embedded applications from the \textit{Embench} benchmark suite. When synthesized, RISSPs can achieve 8-to-43\% reduction in area and 3-to-30\% reduction in power compared to a processor supporting the full RISC-V ISA, and are also on average \textasciitilde40 times more energy efficient than \textit{Serv} - the world's smallest 32-bit RISC-V processor. When physically implemented as FlexICs, the three extreme edge RISSPs achieve up to 42\% area and 21\% power savings with respect to the full RISC-V processor.
\end{abstract}


\input{src/introduction.tex}
\input{src/related_work.tex}
\input{src/methodology.tex}
\input{src/evaluation.tex}
\input{src/compiler.tex}
\input{src/conclusions.tex}

\section{Acknowledgments}
   This work is partially funded by GRASP EU Project No 101134936. The authors would like to thank John Biggs, a co-founder of Arm and a Pragmatic fellow, for his valuable inputs on physical implementation analysis, and also Bruce Hoult from the RISCV sub-reddit forum for his comments on macro retargeting.

\bibliographystyle{plain}
\bibliography{references}

\end{document}

%% file: src/introduction.tex
\section{Introduction}
\label{sec:intro}

A new class of emerging applications requires embedded electronics capable of sensing, computing, and communicating while also offering qualitative features such as low cost, conformability, comfort, biocompatibility, and sustainability in addition to low power consumption. This new class of applications is now described as “Extreme Edge” which extends beyond edge computing today \cite {eenews}.

Extreme edge applications can be classified into two main categories: 1) \textbf{Short-lived} applications that have short lifetimes. Item-level intelligence, Fast Moving Consumer Goods (FMCG) and single-use healthcare products are examples of short-lived applications. Since electronics are part of the product itself, they will have a short lifetime, and any firmware/software update to the electronics will not be very likely. For instance, a smart dressing product has built-in electronics to monitor wound healing conditions. The lifetime of the dressing is in the range of days before it is replaced. The dressing and its built-in electronics will be disposed. 2) \textbf{Long-lasting} applications that have much longer lifespan. Smart textile, wearable healthcare and environmental/agriculture monitoring are examples of long-lasting extreme edge applications. For instance, the electronics embedded into a smart garment can be monitoring the health status and performance of a person. The garment will be rewashed and reused in a lifespan of several months before it wears out, and its built-in electronics may receive software/firmware updates during its lifetime. \textbf{Table \ref{tab:classification}} summarizes the application domains and features of these two categories.

\begin{table}[!h]
    \centering
    \caption{Classification of extreme edge applications, application domains and distinct characteristics}
    \label{tab:classification}
    \newcolumntype{M}[1]{>{\centering\arraybackslash}m{#1}}
    \bigskip
    \begin{tabular}{|M{5.0cm}|M{5.0cm}|M{5.0cm}|} \hline
    \textbf{Category} & \textbf{Short-lived} & \textbf{Long-lasting} \\ \hline
    \textbf{Domains}          &  Item-level intelligence (e.g., smart labels), FMCG (e.g., smart food/drink packaging), Single-use healthcare (e.g., patches, wound dressings) & Smart textile (e.g., smart garment and rugs), Wearable healthcare (e.g., implantables and ingestibles), Environmental/agriculture monitoring (e.g., smart plant patches) \\ \hline
     \textbf{Required Qualitative Features}  &  Ultra Low-cost,  Conformability, Patient Comfort and Sustainability &  Low-cost, Conformability, Biocompatibility and Sustainability \\ \hline
    \textbf{Application Lifespan}        &  Days/Weeks &  Months/Years \\ \hline
    \textbf{Requirement of Software and Firmware Updates}  &  Not likely &  Very likely \\ \hline
   
    \end{tabular}
\end{table}

Conventional silicon-based electronics are unable to meet the requirements of extreme edge applications because they do not satisfy many of the required features. For example, smart labels in asset tracking and smart packages in FMCG are ultra cost-sensitive applications where electronics embedded into them must not cost more than a few cents per unit to ensure economic viability. An electrocardiogram (ECG) patch requires a comfortable and conformable form factor in electronics for the patient where conventional bulky electronics will cause discomfort. An implantable or ingestible device must contain electronics that are biocompatible and conformable. The volume of extreme edge products in terms of units is enormous, measured in a few trillions annually \cite{b41,b42,b43,b47}. If electronic devices are embedded into such volume of products, they must be manufactured in sustainable semiconductor fabs that have a low environmental footprint.

\begin{figure}[!h]
    \centering
    \includegraphics[width =0.6\columnwidth]{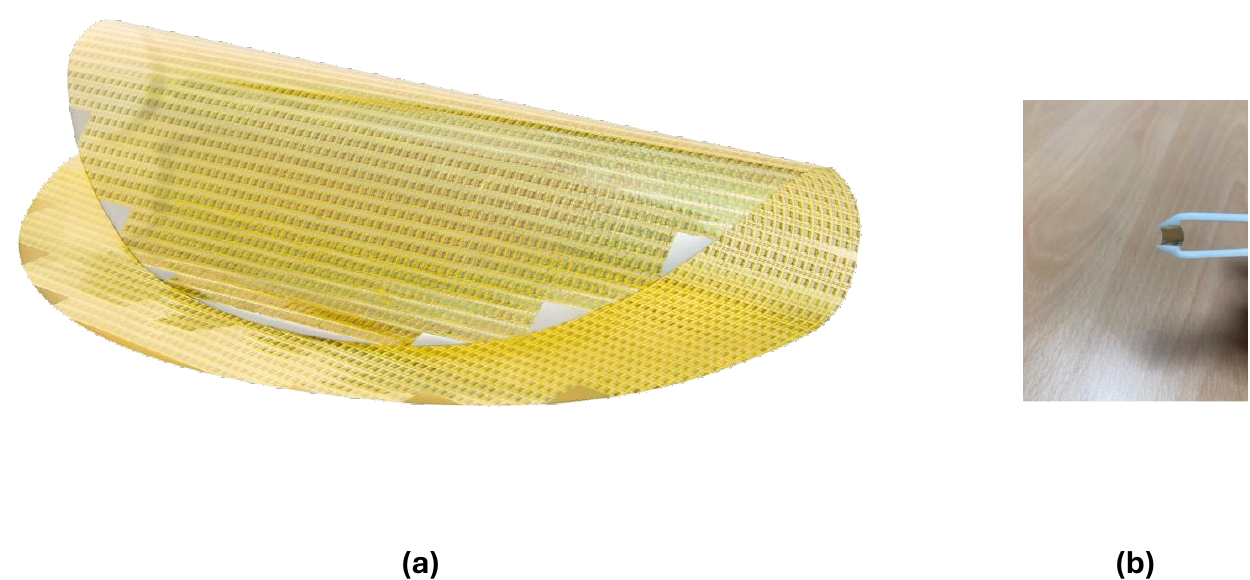}
    \caption{FlexIC technology. (a) A 200 mm wafer on a polyimide substrate, (b) A FlexIC is an integrated circuit manufactured on a polyimide wafer substrate. The semiconductor material is metal-oxide called Indium-Gallium-Zinc-Oxide (IGZO) that can be manufactured at low temperatures. The image of the FlexIC diced from the polyimide wafer is shown. It is an ultra-thin (30 ~\textmu m) and physically flexible or bendable.}
    \label{fig:flexic}
\end{figure}

Flexible electronics \cite{b12} is an alternative technology that meets many of the qualities required by extreme edge. The aim is to produce electronic components that include sensors, compute engines, communication interfaces, batteries using low-cost printing, and lithography techniques on flexible substrates (e.g., plastic or paper). For example, flexible integrated circuits (FlexICs)  \cite{b11} can be fabricated using metal-oxide thin-film transistor technology on a polyimide wafer, as shown in \textbf{Figure \ref{fig:flexic}}. 

The chips are ultra-thin, conformable, bendable and ultra-low cost \cite{b14}, and the semiconductor fabs that manufacture FlexICs have significantly lower carbon footprint per FlexIC than an equivalent Si chip \cite{ahamed24}. Because the fabrication process is low-cost and fast, the development of custom FlexICs specifically designed for an application or domain can be faster and lower in cost than Silicon technology.

The performance requirements of the two extreme edge categories are not demanding where a sensor or an array of sensors are connected to an analogue frontend to pre-process sensor raw signals that are converted into digital data and handled by a digital backend to translate them into knowledge. The knowledge is stored and/or directly transmitted to the external world through a communication interface (e.g., RF). Data sampling rates required by sensors are 200 Hz\cite{b32}, which implies that digital backend processing speeds can be in the range of Hz for many extreme edge applications or kHz for applications that require faster response.

Natively flexible smart systems \cite{b13} consisting of flexible electronic components, such as sensors, displays and batteries coupled with compute engines, memory and communication units, are a natural fit for extreme edge. Both general-purpose processors and domain-specific processing engines have been developed as FlexICs. For instance, the studies in \cite{b18, b19, b20, b21, b14, bleier22} demonstrated natively flexible 4-bit, 8-bit and 32-bit general-purpose processors whilst others proposed natively flexible domain-specific processing engines in the domains of RFID communication \cite{b30, b33}, machine learning \cite{b22, b23, b24, b25, b26, b28}, security \cite{b32}, and healthcare \cite{b27, b29}. These natively flexible processors have been shown to operate at the kHz range reaching up to 100 kHz, which is sufficient to meet the performance requirements of extreme edge applications.


Although designing a domain-specific processing engine is more efficient than a general-purpose processor, it takes a considerable time to design, test and verify the design if the ASIC approach is taken for domain-specific processing. An alternative is to use an application-specific instruction processor (ASIP) \cite{b34, b35}, where the instruction sets are customized based on the application or domain. However, this approach needs to generate not only the ASIP but also the associated software toolset, such as a compiler, linker and simulator.

In this paper, we propose a methodology for automatically generating domain-specific processors for extreme edge applications. The proposed methodology relies on the open standard RISC-V \cite{b31} instruction set architecture (ISA) and generates processors that use only a subset of the RISC-V full ISA (e.g., RV32I). We observe (see \textit{Section \ref{sec:evaluation}}) that extreme edge and extreme edge-like embedded applications only use 24-86\% of the full RISC-V ISA. Thus, a processor supporting a subset of the full RISC-V ISA can be more efficient in terms of performance, power and area (PPA) because the logic related to the unused instructions is avoided. We define them as \textbf{R}ISC-V \textbf{I}nstruction \textbf{S}ub\textbf{S}et \textbf{P}rocessors or \textbf{RISSP}s.

\begin{figure*}[!ht]
    \centering
    \includegraphics[width = .95\textwidth]{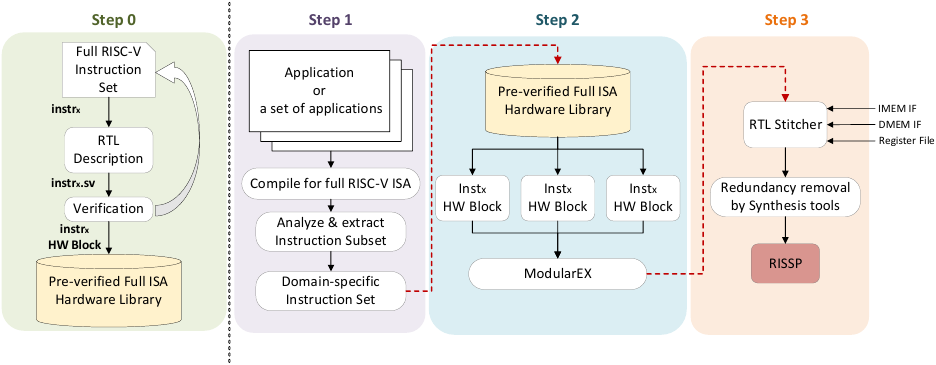}
    \caption{RISSP generation methodology. STEP 0: Development of the pre-verified full ISA hardware library, which is the one-time development effort of the methodology; STEP 1: Generation of domain-specific instruction subset from an application or a set of applications from a domain; STEP 2: Pulling instruction hardware blocks for the domain-specific instruction set from the pre-verified full ISA hardware library, and formation of ModularEX; STEP 3: Construction of the RISSP by stitching ModularEX with fixed units like memory interfaces and the register file}
    \label{fig:methodology}
\end{figure*}

The RISSP generation methodology takes a very different approach to a processor microarchitecture where verification is an integral part of the design process. The main concept is to design a RISSP from pre-verified building blocks rather than the traditional design-followed-by-verification approach.  First, the semantics of all instructions in the ISA are written in RTL as \textbf{instruction hardware blocks}. Then, they are formally verified, and stored into a library. A RISSP is formed by extracting the required instruction hardware blocks from the library, composing them in a modular manner and reducing the verification effort significantly. Since extreme edge applications do not require high-speed operating clock frequencies, single-cycle non-pipelined RISSPs suffice to meet their performance requirements.

The RISSP generation methodology avoids the significant effort of generating proprietary software toolsets specific to ASIPs by relying on the RISC-V software ecosystem. For short-lived extreme edge applications, we predict that RISSPs do not need software updates due to the short life of electronics embedded in the extreme edge device. For long-lasting extreme edge applications, occasional software updates to RISSPs are needed. In this context, the updated application(s) need to be recompiled but the generated code must be constrained to the instruction subset supported by the target RISSP. This requires several modifications in the RISC-V compiler (e.g., gcc) backend to constrain code generation to the instruction subset of the RISSP. However, these modifications involve various challenges at multiple levels of the compiler stack. Instead, we develop a Generative AI-based tool that receives the assembly code of the updated application(s) recompiled to the full ISA, and rewrites the new instructions that are not supported by the RISSP in terms of the instruction subset.

The proposed methodology can automatically generate RISSPs that are 8-to-43\% smaller and 3-to-30\% lower power, when synthesized, compared to a processor supporting the full RISC-V ISA, also generated by the same methodology, and on average about 40 times more energy efficient than \textit{Serv} \cite{b39}. When physically implemented as FlexICs, the extreme edge RISSPs achieve up to 42\% area and 21\% power savings relative to the full RISC-V processor.

The contributions of this paper are summarized as follows:
\begin{enumerate}
\item Automatic generation of RISC-V instruction subset processors (RISSPs) that implement only the subset of the RISC-V ISA targeting extreme edge applications, whilst still relying on the existing software ecosystem
\item Design of a RISSP using a pre-verified instruction hardware block library to significantly reduce the verification time and time-to-market of RISSPs
\item A Generative AI based tool to retarget the new instructions in the updated application(s) to the instructions in the instruction subset supported by RISSP for long-lasting extreme edge applications
\end{enumerate}

The remainder of this paper is structured as follows:\textit{ Section \ref{sec:related}} discusses the related work in processor customization.\textit{ Section \ref{sec:methodology}} presents the proposed methodology of generating RISSPs. \textit{Section \ref{sec:evaluation}} describes the experimental evaluation and also presents the physical implementation results of RISSPs as FlexICs. \textit{Section \ref{sec:recompilation}} introduces the Generative AI-based code recompilation/retarget framework. Finally, \textit{Section \ref{sec:conclusions_and_fw}} concludes the paper.

%% file: src/related_work.tex
\section{Related Work}
\label{sec:related}

There are two main approaches in the design of domain-specific processors: i) Top-down or subtractive customization approach where the customization is performed over an already designed processor by removing the gates that are not used by the target application; ii) Bottom-up or additive approach where the customization is performed at the design level by placing in the hardware blocks required by the applications. 

PDAT \cite{b1} and Bespoke processors \cite{b2} are two examples of a subtractive customization approach. PDAT converts a baseline processor into a standard netlist and uses property assertions to determine whether a specific gate is utilized during the execution of an embedded application. It exploits gate-level property checking to replace unused cells with wires if the relevant property is not asserted. Bespoke processors focus on the development of processing hardware from a baseline processor customized to a specific application. It uses input-independent gate activity analysis to remove the gates from the baseline processor that do not toggle during execution, resulting in a bespoke processor. The main disadvantage of the subtractive approach is that the opportunity of finding and removing a large number of unused gates and hardware blocks in a baseline processor is limited because processors are typically designed as lean as possible. For example, an arithmetic logic unit (ALU) is shared among many instructions, and these subtractive approaches are conservative to remove the portions of the ALU even if some of the ALU-using instructions can be removed.

ASIPs \cite{b3, b4, b34, b35} use an additive customization approach in which they create the custom instruction set for the target application or domain and then generate the ASIP that executes the custom instruction set. A complete software toolchain (e.g., a compiler, assembler, linker and simulator) is also generated to support the ASIP, which is an overhead. They adopt conventional processor microarchitecture techniques, and verification is not an integral part of the design. Conventional verification strategies must be used after generating ASIPs.

On the other hand, verification is an integral part of the RISSP generation methodology, and RISSPs are created in a bottom-up approach using the novel pre-verified instruction hardware block library concept reducing the verification effort of the generated custom processor. In addition, RISSPs rely on open-source software toolchain of the RISC-V software ecosystem rather than producing a new software toolchain for a proprietary ISA like in ASIPs.

%% file: src/methodology.tex
\section{Methodology of RISSP Generation}
\label{sec:methodology}

RISC-V \cite{b31} is one of the most popular ISAs that is rapidly established in various embedded domains. A key design principle of the RISC-V ISA is its modularity allowing computer architects to design custom processors based on the requirements of the target application. Additionally, the built-in support for custom instructions offers a significant advantage compared to other ISAs. While a modular ISA allows independent instruction groups to be separated and combined based on the application requirements, the modularity of the ISA implementation depends on how these instruction groups or extensions are designed in hardware. It is important to note that a modular ISA does not necessarily infer that the underlying implementation of the ISA is modular. Unless the ISA implementation inherently supports ``instruction-level modularity'', the only way to remove a subset of the ISA is to modify the actual RTL design. However, removing support for a subset of the ISA from the actual ISA implementation is not straightforward and requires a deep understanding of the RTL implementation of the processor supporting the ISA. This process may introduce errors in the design and extensive verification is necessary to ensure that the modifications maintain the functionality. 

One of the key benefits of this work is to enhance the instruction-level modularity of the ISA implementation by defining instructions as independent, fully functional, pre-verified RTL blocks.

\begin{table}[!h]
    \centering
    \caption{A high-level overview of instruction hardware blocks of the RISC-V RV32I/E ISA}
    \label{tab:insn_blocks}
    \renewcommand{\arraystretch}{1.0}
    \newcolumntype{M}[1]{>{\centering\arraybackslash}m{#1}}
    \bigskip
    \begin{tabular}{|M{1.5cm}|M{3.5cm}|M{5.8cm}|} \hline
    \textbf{Type}    & \textbf{Instructions} & \textbf{Block Diagram} \\ \hline
    B-type           & beq, bne, blt, bge, bltu, bgeu                                                   & \includegraphics[width = .3\columnwidth]{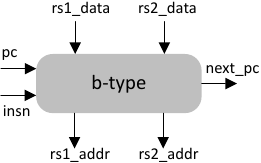} \\ \hline
    R-type           & add, sub, sll, slt, sltu, xor, srl, sra, or, and                                 & \includegraphics[width = .3\columnwidth]{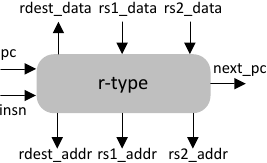} \\ \hline
    I-type           & lb, lh, lw, lbu, lhu, addi, slli, slti, sltiu, xori, srli, srai, ori, andi, jalr & \includegraphics[width = .3\columnwidth]{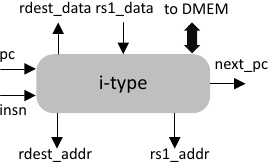} \\ \hline
    S-type           & sb, sh, sw                                                                       & \includegraphics[width = .3\columnwidth]{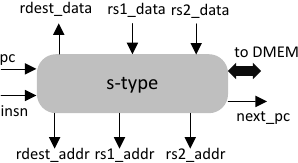} \\ \hline
    U-type           & lui, auipc                                                                       & \includegraphics[width = .25\columnwidth]{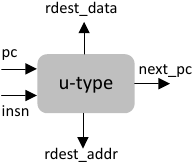}  \\ \hline
    J-type           & jal                                                                              & \includegraphics[width = .25 \columnwidth]{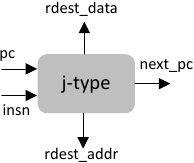}  \\ \hline
    \end{tabular}
\end{table}

\subsection{Concept of Instruction Hardware Blocks}
The RISSP generation methodology relies on the RISC-V ISA and its software toolchain. A RISSP is automatically generated for a given application or a domain of similar applications, as shown in \textbf{Figure \ref{fig:methodology}}. It starts with the development of the pre-verified full ISA hardware library (see Step 0). This is a one-time effort for a given ISA. The pre-verified full ISA hardware library contains ``formally verified'' instruction hardware blocks, and is, in concept, similar to the standard cell library designed for a process technology used in chip design and implementation, which will be described in \textit{Section \ref{subsubsec:fullISA_lib}}. Next, we characterize the application or the domain of applications by compiling them to the target RISC-V ISA (in our case RV32E) to identify the domain-specific instruction subset (see Step 1). The hardware description of each instruction in the instruction subset is called ``instruction hardware block'' written in RTL (i.e., SystemVerilog), and extracted from a pre-verified full ISA hardware library of all instructions in RISC-V ISA.

Each instruction in the RISC-V ISA can be thought of as a discrete, fully functional hardware block. \textbf{Table \ref{tab:insn_blocks}} highlights the block diagram of each instruction type in the ISA. The semantics of each instruction is converted into its hardware description in RTL, along with a glue logic interfacing with the Register File (RF) and the memory. The extracted instruction hardware blocks are then automatically combined to build the Modular Execution Unit (ModularEX).

\subsection{Modular Execution Unit}
ModularEX is the main execution unit of the generated RISSP. It consists of a switch that handles the instruction hardware blocks and their interfaces. The switch is automatically generated and ensures that the outputs of the instruction-under-execution module are triggered. It is a simple case statement in SystemVerilog with N cases, where N is the number of the instructions in the sub-ISA. The instruction hardware blocks are integrated with the switch into ModularEX (fully combinational design).


In Step 3 - \textbf{Figure \ref{fig:methodology}}, ModularEX is stitched to the memory interfaces and the RF to generate an unoptimised RISSP, which is an input to the hardware synthesis tool that optimizes it by maximizing sharing inside ModularEX. The output is the final RISSP tailored to the application or the domain under consideration, which is fully programmable within the boundaries of its instruction subset.

\subsection{RISSP Construction}
\label{subsec:RISSP_construction}
\begin{figure*}[!h]
    \centering
    \includegraphics[width = .85\textwidth]{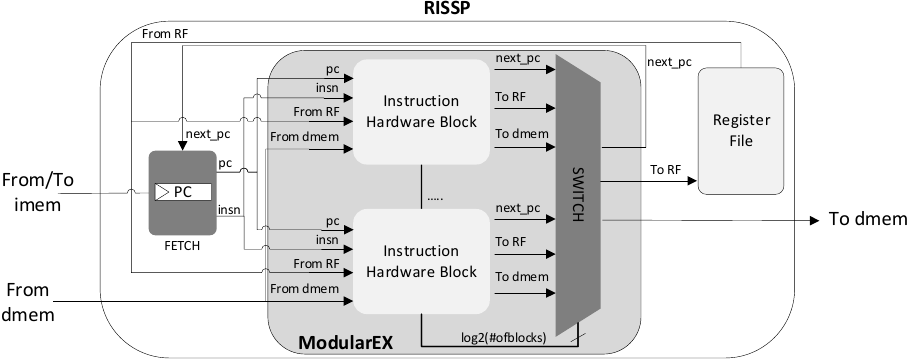}
    \caption{A single-cycle RISSP microarchitecture created by stitching together ModularEX, the fixed units (i.e., Fetch and RF) and Memory Interfaces}
    \label{fig:RISSP}
\end{figure*}

\textbf{Figure \ref{fig:RISSP}} shows the construction of a RISSP by stitching ModularEX with the rest of the fixed units such as the fetch unit and the RF as well as memory interfaces. By construction, a RISSP has a single-cycle processor microarchitecture where an instruction is fetched from the instruction memory using a 32-bit Program Counter (PC), and sent to the ModularEX unit. Then, the switch selects the actual instruction hardware block of the fetched instruction executing it at the same cycle. In other words, the switch performs as a partial decoder that selects which block will be enabled at every clock cycle. The full decoding of the instruction is performed inside each instruction block.

The key benefit of the RISSP generation methodology is that it does not require any expertise in microarchitecture, RTL design or gate netlists. The only input of the framework is the domain-specific instruction subset identified by the application or the domain. All design optimization decisions are taken by  the synthesis tool rather than optimizing the design manually. This is a conscious decision because contemporary commercial/non-commercial hardware synthesis tools are very efficient in discovering redundancies and eliminating them. For instance, the synthesis tool will optimize the gate netlists by maximizing the resource sharing if multiple instruction hardware blocks have common operations among them (e.g., design flattening). Any logic optimization performed during synthesis will be formally verified using equivalence checking by the synthesis tool. This approach may not lead to "minimal area/power" but minimizes the design and verification time.

\subsection{Verification}
\label{subsec:verifcation}
Verification is a critical and complex phase in processor design. The fundamental benefit of the proposed methodology is the ability to streamline and simplify the verification process.

\subsubsection{Development of the Pre-verified Full ISA Hardware Library}
\label{subsubsec:fullISA_lib}
Generating a RISSP using the novel concept of a pre-verified ISA hardware library significantly reduces the verification complexity of RISSP. The full ISA hardware library has been extensively verified in multiple ways including formal verification. The development of the pre-verified ISA hardware library is a one-time verification effort that becomes the non-recurring-engineering (NRE) cost of RISSP. The NRE is amortized by the number of RISSPs shipped during the lifetime of this library. In contrast, any new design in conventional CPU/ASIP methodologies requires design and verification, i.e., verification effort becomes a recurring cost.

\begin{figure*}[!ht]
    \centering
    \includegraphics[width = \textwidth]{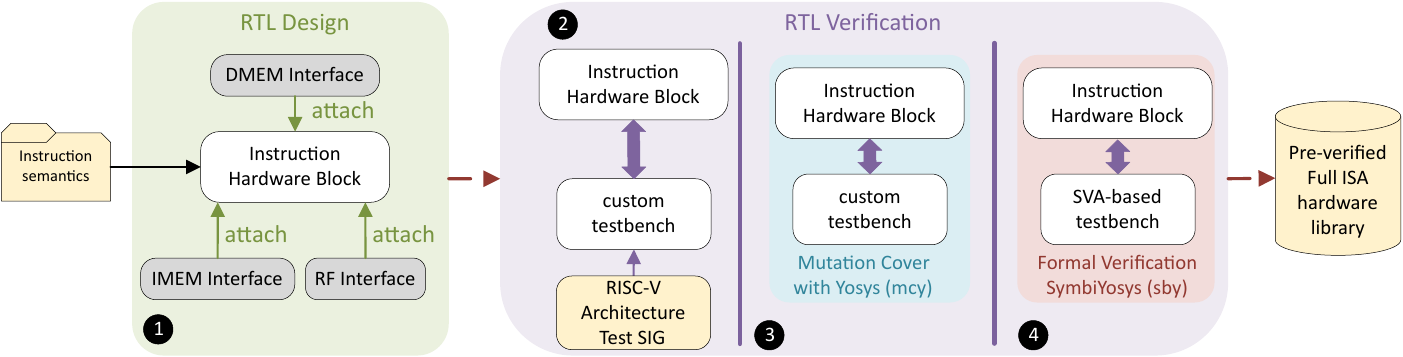}
    \caption{The development of the pre-verified full ISA hardware library}
    \label{fig:hardware_lib}
\end{figure*}

The full ISA hardware library is a collection of instruction hardware blocks written in SystemVerilog. The library supports the RISC-V RV32I/E instruction set and is fully extendable to support other groups of RISC-V instructions or even custom instructions. Each instruction hardware block is formally verified against its specification in the ISA. This is a significant step towards instruction modularity in the RISSP design and its simplified verification. One of the key advantages of this approach is that verification of instruction hardware blocks is integrated into the RISSP development process. Every instruction hardware block is formally pre-verified before it is used for the generation of a RISSP.

\textbf{Figure \ref{fig:hardware_lib}} presents the development process of the pre-verified full ISA RTL library. Starting with the RTL design, the instruction semantics are combined with pre-developed standard interfaces - used for communication with the rest of the processor - to define the actual instruction hardware block (\textbf{Figure \ref{fig:hardware_lib}} - \circled{1}). Following the design phase, there are three distinctive verification steps before including the new instruction hardware block in the library. This process includes: (a) the development of a testbench that automatically verifies the instruction hardware block using standard test cases (\textbf{Figure \ref{fig:hardware_lib}} - \circled{2}), (b) self-checking of the testbench to ensure that can verify the instruction hardware block under test (\textbf{Figure \ref{fig:hardware_lib} }- \circled{3}), and (c) the formal verification of the instruction hardware block using SystemVerilog Assertions (SVA) (\textbf{Figure \ref{fig:hardware_lib}} - \circled{4}).

For functional verification, individual test cases from the RISC-V Foundation Architecture Test SIG \cite{b5} are automatically extracted for each instruction hardware block, and custom testbenches have been developed to verify these test cases. Specifically, RISC-V Foundation Architecture Test Suite provides an extensive set of test cases for each instruction, developed in Assembly, ensures that the RISC-V specification has been interpreted correctly and the implementation of the target processor is RISC-V Architecture test compliant. The toolflow automatically isolates the tests cases for each instruction, transforms them in a binary format, and passes the test vector to the instruction hardware block through our custom testbenches. This process iteratively verifies each instruction hardware block for the complete set of test cases.

To further extend our verification framework of the library, each testbench is evaluated with MCY tool from \textit{YosysHQ} suite \cite{b6}. MCY is one of the most popular tools for verification primarily because it offers the ability to generate several mutations of the target RTL design (in our case the implemented instruction blocks). The tool automatically selects only the mutated designs that can cause an important change in the output of the design by filtering them using formal verification techniques. The final step is to automatically check that the testbench can detect and fail for the mutations that can cause an important change in the output of the target design. Under this environment, we ensure that our testbenches are capable of verifying the target instruction hardware blocks.

In terms of formal verification, the full ISA library is formally verified using \textit{SymbiYosys} \cite{b7} (sby), a frontend for \textit{Yosys} for Formal Verification. For every instruction hardware block, a set of SVA are developed to ensure that the instruction semantics are satisfied. Also, additional assertions are used to validate the behavior of the interfaces. This approach simplifies the verification process of a RISSP, as each instruction hardware block is already independently and formally verified. Moreover, the process benefits from the well-defined semantics of each instruction in the ISA. 

Instruction hardware block representations of all instructions in the ISA are functionally and formally verified and are stored in the pre-verified full ISA hardware library. When constructing a RISSP, instruction hardware blocks from the library will be stitched together to build the ModularEX that does not require any further verification. Therefore, the pre-verified full ISA hardware library concept is similar to the concept of standard cell libraries in VLSI design where the standard cell library for a foundry process is created once and re-used in many chip designs. Each cell is a fully verified building block in a finer granularity.

\subsubsection{Integration-level Verification}
\label{subsubsec:integver}

The pre-verified ModularEX is integrated into the pre-verified fixed units, such as fetch and register file to construct the RISSP. Various verification processes and unit-tests have been developed to ensure the correctness of the integrated RISSP. They include custom domain checks at various points, such as checks at instruction and register level to verify that each instruction is fetched and executed correctly. Additionally, RISC-V compatibility is tested with \textit{RISCOF}, an architectural test framework \cite{b8}. \textit{RISCOF} automatically chooses a set of test cases for the RV32E ISA, the RISSP executes these test cases, and writes a signature to a file. Then, this signature is compared with a signature of a reference model that has been obtained from a RISC-V simulator, such as \textit{Spike} \cite{b40}. 

In terms of formal verification at processor-level, RISSP has been verified up to a specific depth with \texttt{riscv-formal} \cite{b49}. This framework is an open-source standard for the formal verification of RISC-V processors and is based on the RISC-V Formal Interface (RVFI), a standard interface for processors to communicate and verification tools. Specifically, RISSP implements the RVFI interface that mainly tracks valid instructions, register file data and addresses, PC changes and memory accesses. With \texttt{riscv-formal}, we perform various and extensive checks at instruction level for correct execution, checks to ensure that the registers maintain consistency, and PC checks that verify the correct PC values before and after the execution of an instruction.

%% file: src/evaluation.tex
\begin{figure*}[!h]
    \centering
    \includegraphics[width =1.0\textwidth]{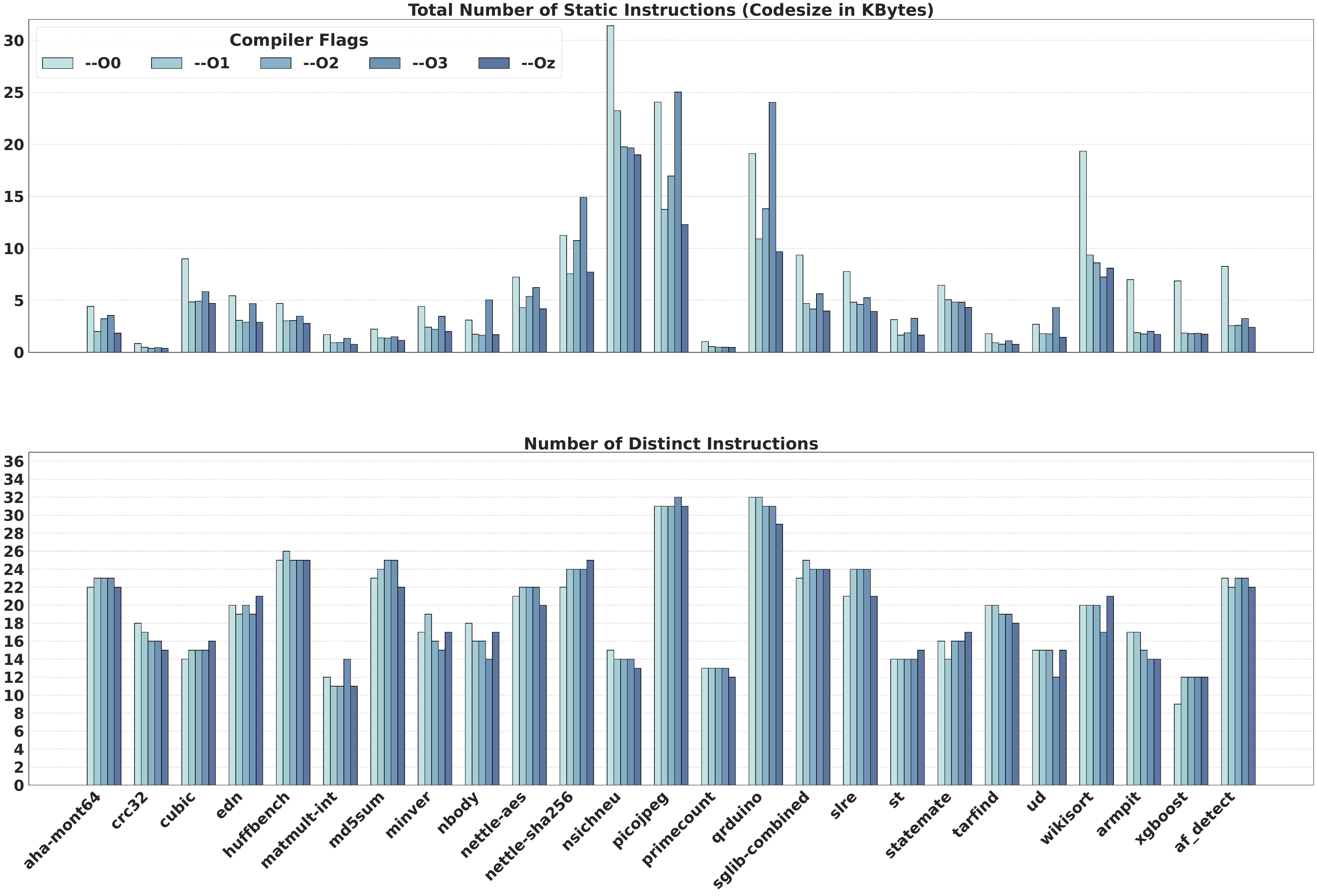}
    \caption{Results of instruction profiling and characterization of the three extreme edge applications, and the benchmarks in \textit{Embench} using different compiler optimization flags. The top diagram presents the codesize in KBytes and the bottom diagram illustrates the number of distinct instructions from the RV32E ISA across different optimization flags.}
    \label{graph:bench_analysis}
\end{figure*}

\section{Experimental Evaluation}
\label{sec:evaluation}
One of the key challenges in evaluating the RISSP generation methodology is the lack of a representative benchmark suite for extreme edge. We select two extreme edge applications previously demonstrated in the FlexIC technology \cite{b25} \cite{b27}, and also an ML kernel \textit{XGBoost} (eXtreme Gradient Boosting) \cite{b36} as one of the key ML algorithms used in many classification tasks as an extreme edge kernel.

\par{\textit{Extreme edge application 1:}} Armpit malodour classification  or \textit{armpit} \cite{b25} is a real-world application in the domain of deodorant industry that advocates the use of flexible organic sensors to detect volatile organic components in odour with a sensor readout interface and machine learning hardware implemented as a custom FlexIC to score malodour in a human armpit. It is a representative example of a natively flexible smart system \cite{b13} consisting of all flexible components from sensors to processing. Two decision trees (one for each gender) are implemented in hardware to classify armpit malodour. The decision trees in the armpit application are used as the benchmark in this paper. The C code of the decision trees was compiled to RV32E to generate a RISSP for the application.

\par{\textit{Extreme edge application 2:}} Atrial fibrillation (AF) detection or \textit{af\_detect} \cite{b27} is another real-world application in the domain of ECG that uses flexible electronics to detect AF events in a wearable ECG patch. An AF prediction algorithm called ``Approximate Pair Presence Tracking'' or APPT was implemented as a custom FlexIC. APPT has three stages: i) detection of R peaks in the ECG signal; ii) calculation of RR intervals (the interval between two R peaks) and delta of RR intervals; iii) a Bloom filter-based binary predictor (AF versus non-AF) using a RR and delta RR interval map. The APPT algorithm in \textit{af\_detect} is used as the benchmark in this paper, which is compiled for RV32E to generate a RISSP.

\par{\textit{Extreme edge kernel}}: \textit{XGBoost} \cite{b36} is known for its performance in classification and regression problems. It is a more accurate and scalable implementation of gradient boosting. For datasets and problems with simpler requirements, decision trees can be mapped on edge devices in the form of C/C++ or on custom hardware implementation processing units to perform ML classification problems, e.g., \textit{armpit}. We use \textit{XGBoost} to extract a decision tree from a dataset and then implement the generated decision tree in C/C++ to evaluate its RISSP requirements. It can be considered as a kernel in extreme edge applications rather than a full application itself. For evaluation purposes, we use the pima indian diabetes database \cite{b37}, a binary-valued data set that predicts whether a patient shows signs of diabetes. The dataset consists of 768 instances and 8 attributes.

We also select \textit{Embench} that is an open-source benchmark suite for embedded systems. It is designed to support a wide range of embedded systems and consists of 22   C-written applications. Although it may not be perfectly aligned with the requirements of extreme edge, it offers a reasonable and structured approach to assess our methodology. The system requirements for these applications require at least 64 KB of ROM and RAM. However, an operating system and output stream are not required.

\begin{table*}[!ht]
    \centering
    \caption{List of distinct instructions per application when compiled with -O2 option}
    \label{tab:distinct}
    \renewcommand{\arraystretch}{1.0}
    \newcolumntype{M}[1]{>{\centering\arraybackslash}m{#1}}
    \bigskip
    \begin{tabular}{|M{2cm}|M{13cm}|} \hline
    \textbf{Application} & \textbf{List of Distinct Instructions} \\ \hline
        aha-mont64 & [add, addi, and, andi, beq, bge, bgeu, bltu, bne, jal, jalr, lui, lw, or, slli, sltiu, sltu, srai, srli, sub, sw, xor, xori] \\ \hline
        crc32 & [add, addi, andi, bge, bne, jal, jalr, lui, lw, slli, sltiu, srli, sub, sw, xor, xori] \\ \hline
        cubic & [addi, and, andi, beq, bge, blt, bne, jal, jalr, lui, lw, slti, sltiu, sw, xor] \\ \hline
        edn & [add, addi, andi, beq, bge, bne, jal, jalr, lh, lhu, lui, lw, sh, slli, sltiu, sra, srai, srli, sub, sw] \\ \hline
        huffbench & [add, addi, and, andi, beq, bge, bgeu, blt, bltu, bne, jal, jalr, lbu, lui, lw, or, ori, sb, sll, slli, sltiu, srai, srli, sub, sw] \\ \hline
        matmult-int & [add, addi, bge, bne, jal, jalr, lui, lw, slli, sltiu, sw] \\ \hline
        md5sum & [add, addi, and, andi, beq, bge, bgeu, blt, bltu, bne, jal, jalr, lui, lw, or, sb, sll, slli, sltiu, srl, srli, sub, sw, xor, xori] \\ \hline
        minver & [add, addi, and, beq, bge, bne, jal, jalr, lui, lw, slli, slti, sltiu, sub, sw, xor] \\ \hline
        nbody & [add, addi, and, andi, beq, bge, bne, jal, jalr, lui, lw, slli, slti, sltiu, srli, sw] \\ \hline
        nettle-aes & [add, addi, and, andi, beq, bge, bgeu, bltu, bne, jal, jalr, lbu, lui, lw, or, sb, slli, sltiu, srli, sub, sw, xor] \\ \hline
        nettle-sha256 & [add, addi, and, andi, beq, bge, bgeu, bltu, bne, jal, jalr, lbu, lhu, lui, lw, or, sb, slli, sltiu, sltu, srli, sub, sw, xor] \\ \hline
        nsichneu & [add, addi, beq, bge, blt, bne, jal, jalr, lui, lw, slli, sltiu, sub, sw] \\ \hline
        picojpeg & [add, addi, and, andi, beq, bge, bgeu, blt, bltu, bne, jal, jalr, lb, lbu, lh, lhu, lui, lw, or, sb, sh, sll, slli, sltiu, sltu, sra, srai, srli, sub, sw, xori] \\ \hline
        primecount & [add, addi, beq, bge, blt, bne, jal, jalr, lui, lw, slli, sltiu, sw] \\ \hline
        qrduino & [add, addi, and, andi, beq, bge, bgeu, blt, bltu, bne, jal, jalr, lbu, lhu, lui, lw, or, ori, sb, sh, slli, sltiu, sltu, sra, srai, srl, srli, sub, sw, xor, xori] \\ \hline
        sglib-combined & [add, addi, andi, beq, bge, bgeu, blt, bltu, bne, jal, jalr, lbu, lh, lui, lw, sb, sh, slli, sltiu, sltu, srai, sub, sw, xori] \\ \hline
        slre & [add, addi, and, andi, beq, bge, bgeu, blt, bltu, bne, jal, jalr, lbu, lui, lw, or, slli, slt, sltiu, sltu, srai, sub, sw, xori] \\ \hline
        st & [add, addi, and, bge, blt, bne, jal, jalr, lui, lw, slli, slti, sltiu, sw] \\ \hline
        statemate & [addi, beq, bge, blt, bne, jal, jalr, lbu, lui, lw, or, sb, sh, sltiu, sub, sw] \\ \hline
        tarfind & [add, addi, andi, beq, bge, bgeu, bltu, bne, jal, jalr, lbu, lui, lw, sb, slli, sltiu, srli, sub, sw] \\ \hline
        ud & [add, addi, beq, bge, blt, bne, jal, jalr, lui, lw, or, slli, sltiu, sub, sw] \\ \hline
        wikisort & [add, addi, andi, beq, bge, blt, bne, jal, jalr, lui, lw, or, slli, slt, sltiu, sltu, srai, srli, sub, sw] \\ \hline
        armpit & [add, addi, andi, beq, bge, blt, bne, jal, jalr, lbu, lui, lw, slli, sltiu, sw] \\ \hline
        xgboost & [addi, andi, bge, blt, jal, jalr, lui, lw, srli, sw, xor, xori] \\ \hline
        af\_detect & [add, addi, andi, beq, bge, bgeu, blt, bltu, bne, jal, jalr, lbu, lui, lw, sb, sh, slli, sltiu, srai, srli, sub, sw, xor] \\ \hline
    \end{tabular}
\end{table*}

\subsection{Instruction Subset Profile of Applications}
\label{sec:benchmark_analysis}
Before generating a RISSP for each application, we need to characterize each application to identify the distinct instructions from the RV32E instruction set required to implement each RISSP. The next step is to explore how the number of distinct instructions is related to the codesize for different compiler optimization flags.

\textbf{Figure \ref{graph:bench_analysis}} presents the profiling results of the applications from \textit{Embench} and the three extreme edge applications. The applications have been compiled for RV32E instruction set as baremetal (without support of \texttt{stdlib}, \texttt{libc}, \texttt{libgcc} and startfiles) with different compiler optimization flags (\texttt{--O0, --O1, --O2, --O3, --Oz}) using the standard \texttt{riscv32-gnu-toolchain} (version 13.2.0). Through this analysis, we identify the optimal code size with respect to the distinct instructions for each application for the generation of RISSP. The number of distinct instructions ranges from 9 to 32 instructions, and the average is 19 instructions for different optimization flags across all applications. Since the RV32E ISA has around 40 instructions, applications use only 24-86\% of the full ISA. For the extreme edge applications, these numbers are 42\% (\textit{armpit}) 31\% (\textit{xgboost}), and 61\% (\textit{af\_detect}). In terms of total number of static instructions across the different compiler optimization results, the average across all the applications are 2027, 1149, 1207, 1586 and 1018 for \texttt{--O0, --O1, --O2, --O3,} and \texttt{--Oz}, respectively. For the extreme edge applications, we observe a substantial reduction in static instructions, as we progress from \texttt{--O0} to \texttt{--0z}. Specifically, using the \texttt{--O2} optimization level flag, we observe 75\%, 74\% and 69\% less static instructions (reduction in codesize) than \texttt{-O0} for \textit{armpit}, \textit{xgboost} and \textit{af\_detect}, respectively, thus, the difference in the number of distinct instructions between \texttt{--O2} and \texttt{--Oz} is negligible.

These results indicate that there is an opportunity to improve power/area of a 32-bit RISC-V processor by eliminating unnecessary instructions and designing a customized RISSP for a target application. Given that the \texttt{-O2} compiler flag results in minimal discrepancies in average code size with respect to distinct instructions, we select it as the basis for subsequent RISSP design for power, area and frequency analysis. \textbf{Table \ref{tab:distinct}} lists the number of distinct instructions per application when compiled with \texttt{-O2} option.

\subsection{FlexIC Synthesis Results}
\label{sec:power_area_freq}
The RTL of each RISSP is synthesized using \textit{Pragmatic}’s 0.6µm IGZO-based FlexIC process using a commercial EDA tool. Each RISSP is synthesized without the RF to better understand the effects of the instruction subsets in the hardware. We also introduce two baselines to compare the RISSPs. The first baseline is an application-independent RISC-V processor supporting the full RV32E ISA (\textasciitilde40 instructions) generated using the RISSP methodology. This is called \textit{RISSP-RV32E}. The second baseline is \textit{Serv} \cite{b39}, - the world's smallest 32-bit RISC-V processor. \textit{Serv} is a bit serial processor supporting RV32I ISA, and its RF is mapped to on-chip memory rather than having a dedicated RF. \textit{Serv} is configured to support RV32E (i.e., 16 general purpose registers rather than 32).

\begin{figure}[!h]
   \centering
   \includegraphics[width=0.6\columnwidth]{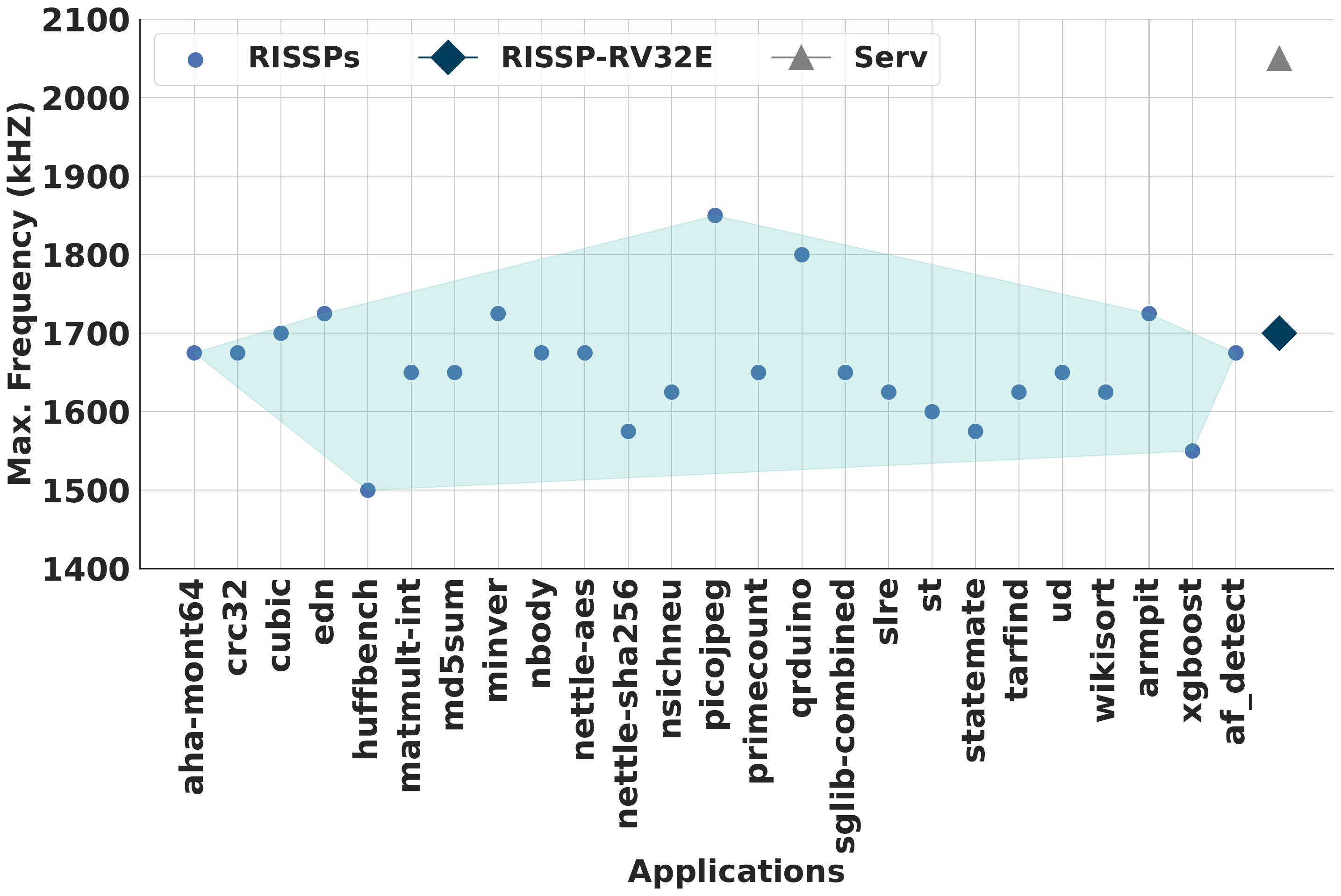}
   \caption{Maximum clock frequencies at which RISSPs, RISSP-RV32E and Serv that can operate}
   \label{graph:freq_analysis}
\end{figure}

\subsubsection{Maximum Frequency Analysis}
\textbf{Figure \ref{graph:freq_analysis}} illustrates the maximum frequency (kHz) of RISSPs compared to the two baselines. To determine the maximum operating frequency, we vary the input clock frequency while adjusting design constraints in the EDA tools. Starting at 100 kHz with significant positive slack, the frequency was incremented by 25 kHz steps until reaching 3 MHz, where the design became over-constrained with negative slack. The highest frequency with positive slack is identified as the maximum.
The maximum clock frequencies of RISSPs range from 1,500 to 1,850 kHz, whilst \textit{RISSP-RV32E} and \textit{Serv} can achieve up to 1,700 kHz and 2,050 kHz, respectively.

\subsubsection{Area Analysis}
\label{sec:area_analysis}
\begin{figure}[!ht]
   \centering
   \includegraphics[width=0.6\columnwidth]{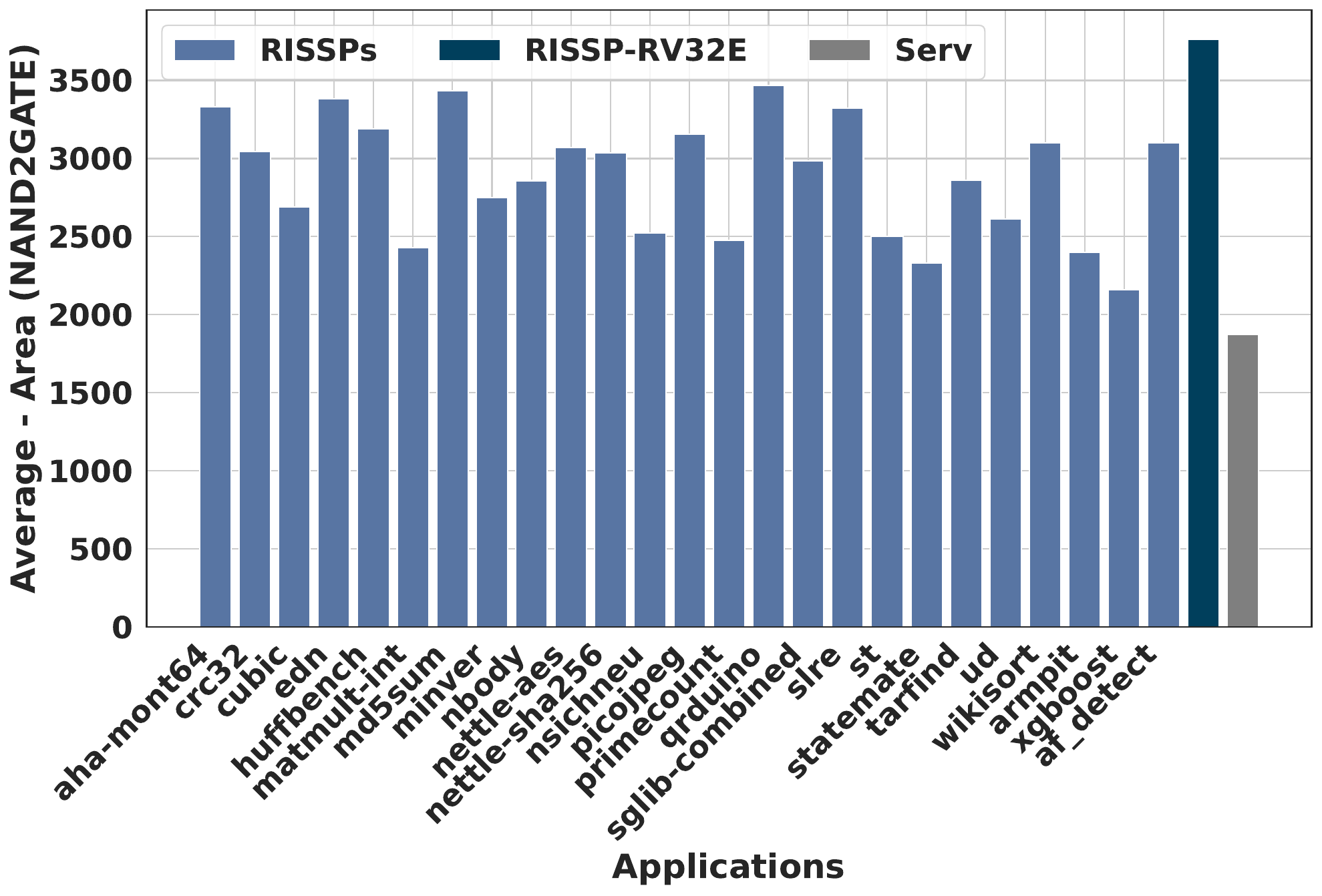}
   \caption{Average NAND2-equivalent gate count of RISSPs compared to Serv and RISSP-RV32E across the range of frequencies with positive slack}
   \label{graph:area_analysis}
\end{figure}

\textbf{Figure \ref{graph:area_analysis}} presents the average area of the RISSP for each application and the baselines. The average area is calculated from the NAND2-equivalent gatecounts of the processor across the range of frequencies with positive slack. RISSPs occupy 8-to-43\% less area compared to \textit{RISSP-RV32E}. The smallest RISSP (i.e., \textit{xgboost}) is 23\% larger than \textit{Serv}.

 \begin{figure}[!ht]
   \centering
   \includegraphics[width =0.6\columnwidth]{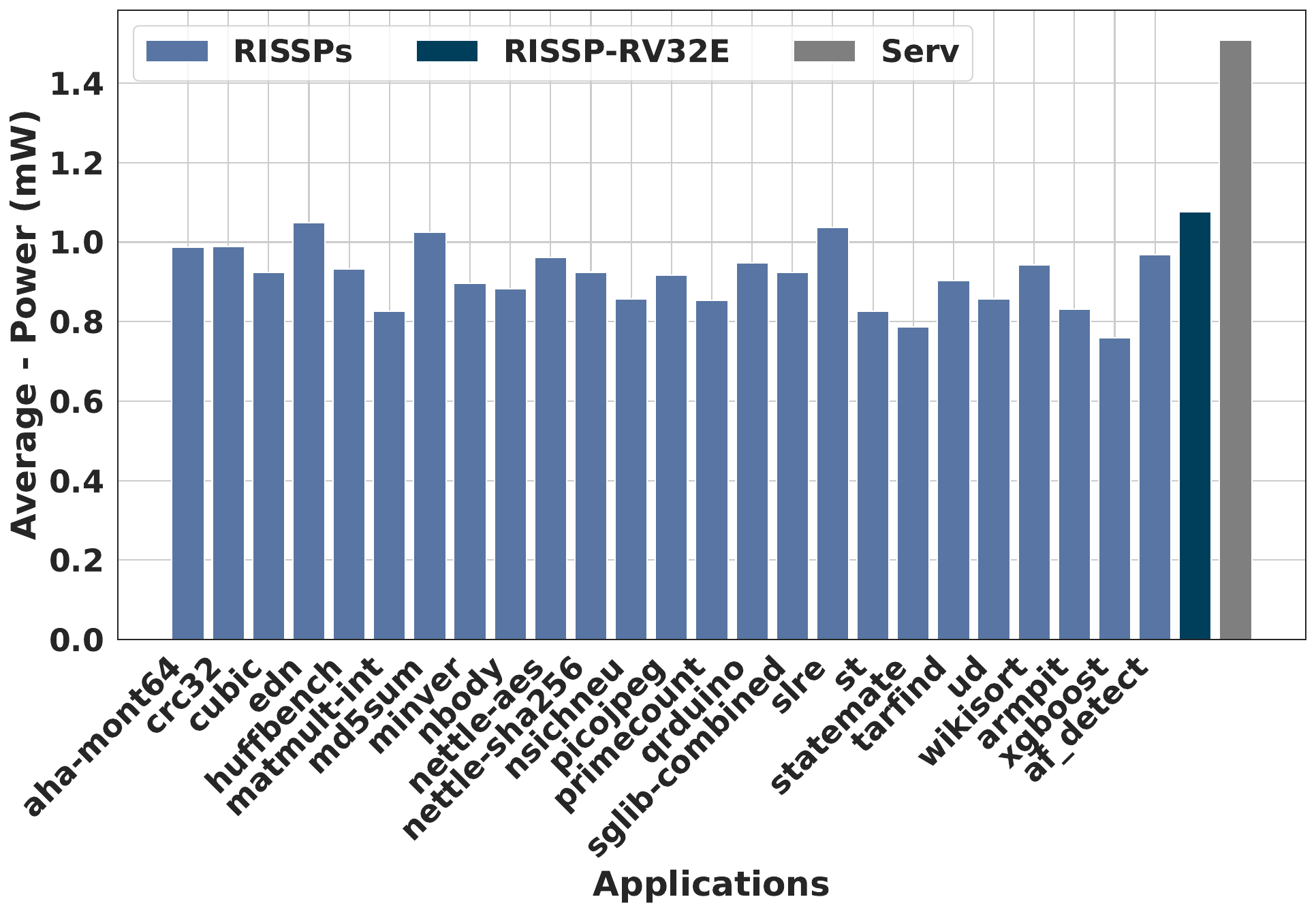}
   \caption{Average power (static + dynamic) comparison of the RISSPs, Serv and RISSP-RV32E across the range of frequencies with positive slack}
   \label{graph:power_analysis}
\end{figure}

\subsubsection{Power Analysis}
 \textbf{Figure \ref{graph:power_analysis}} shows the average power of the RISSP for each application and the baselines. The average powered is calculated in a similar manner to \textit{Section \ref{sec:area_analysis}}. RISSPs consume 3-to-30\% less power than \textit{RISSP-RV32E}. Although \textit{Serv} has a lower gatecount compared to RISSPs and \textit{RISSP-RV32E}, its power consumption is higher than them. For example, it consumes 40\% more power than \textit{RISSP-RV32E}. This is mainly because \textit{Serv} has a higher proportion of flip-flops (FFs) compared to the RISSP microarchitecture, and an FF in the FlexIC process consumes 10 times more power than a NAND2 gate.

\subsubsection{Energy Efficiency}

\begin{figure}[!ht]
    \centering
    \includegraphics[width = 0.6\columnwidth]{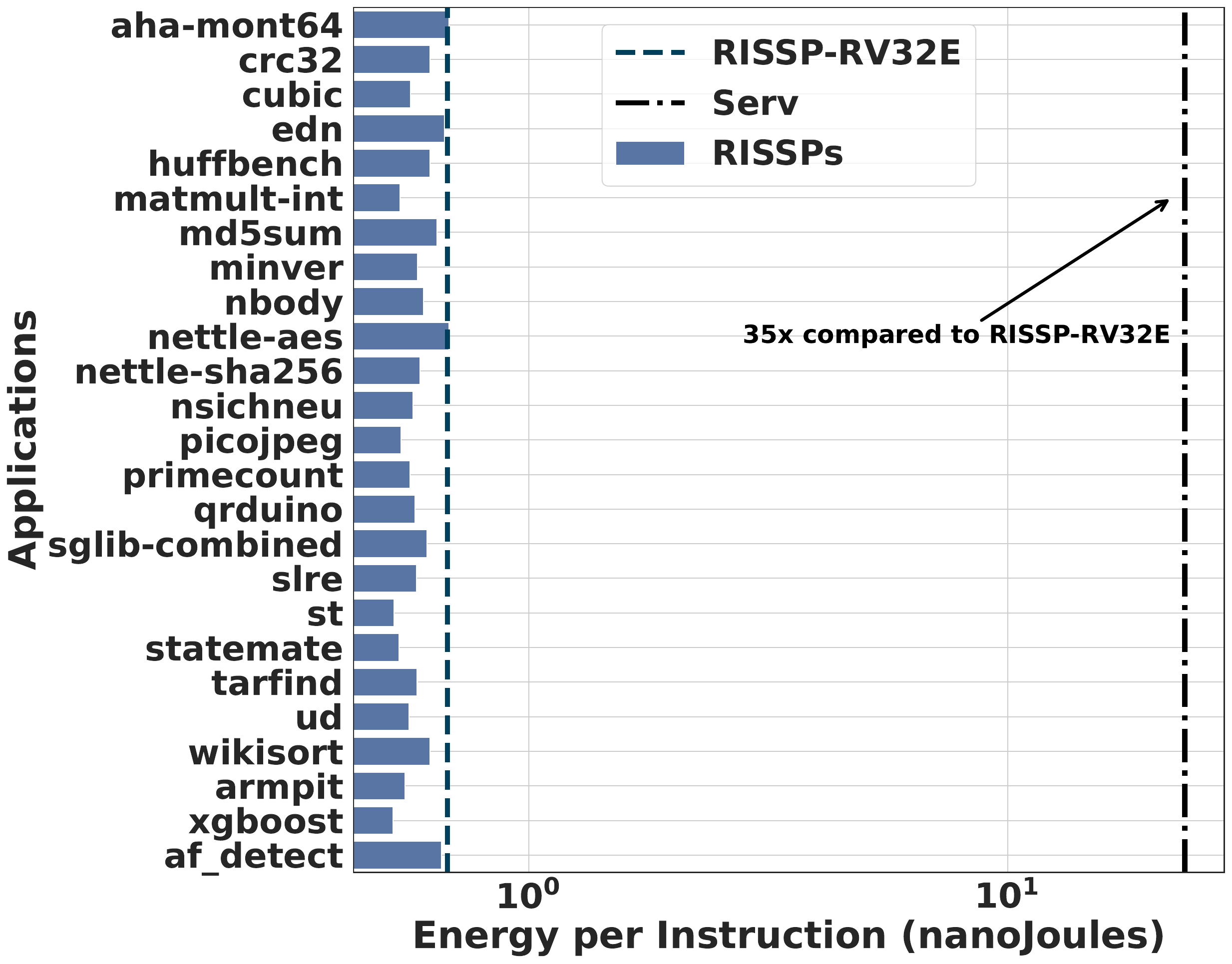}
    \caption{Energy per Instruction (EPI) in nanoJoules for RISSPs and the two baselines (RISSP-RV32E, and Serv)} 
    \label{graph:epi_analysis}
\end{figure}

\textbf{Figure \ref{graph:epi_analysis}} illustrates the Energy per Instruction (EPI) for the RISSPs, \textit{RISSP-RV32E} and \textit{Serv}. The x-axis represents the RISSPs corresponding to each target application. Because a RISSP is a single-cycle processor, the clock cycles per instruction (CPI) for RISSPs and \textit{RISSP-RV32E} is 1. On the other hand, \textit{Serv} is a bit-serial processor, and has a CPI of 32 on average. The EPI is calculated by dividing the power consumption of the design point at the maximum frequency by that frequency and multiply the product by the CPI. The results highlight that \textit{RISSP-RV32E} and RISSPs are about 35 and 40 times more energy-efficient than \textit{Serv}. The bit-serial architecture of \textit{Serv} leads to a high CPI resulting in much higher energy consumption per instruction.

\subsection{FlexIC Physical Implementation Results}
\label{sec:implementation}

We present full physical implementation of the three extreme edge RISSPs as FlexICs and compare them to full physical implementation of the two baselines. RISSPs are generated for the instruction subsets from the applications compiled with the \texttt{--O2} flag. The physical implementation goes beyond hardware synthesis to full-fledged FlexIC layouts by floorplanning, clock tree insertion, and place and route. The goal is to show that the full physical implementation-related overheads (e.g., inserted buffers) do not hide the advantages of the RISSP philosophy (i.e., not supporting unnecessary instructions in the processor is observably beneficial).

\begin{figure*}[!ht]
    \centering
    \includegraphics[width=\textwidth]{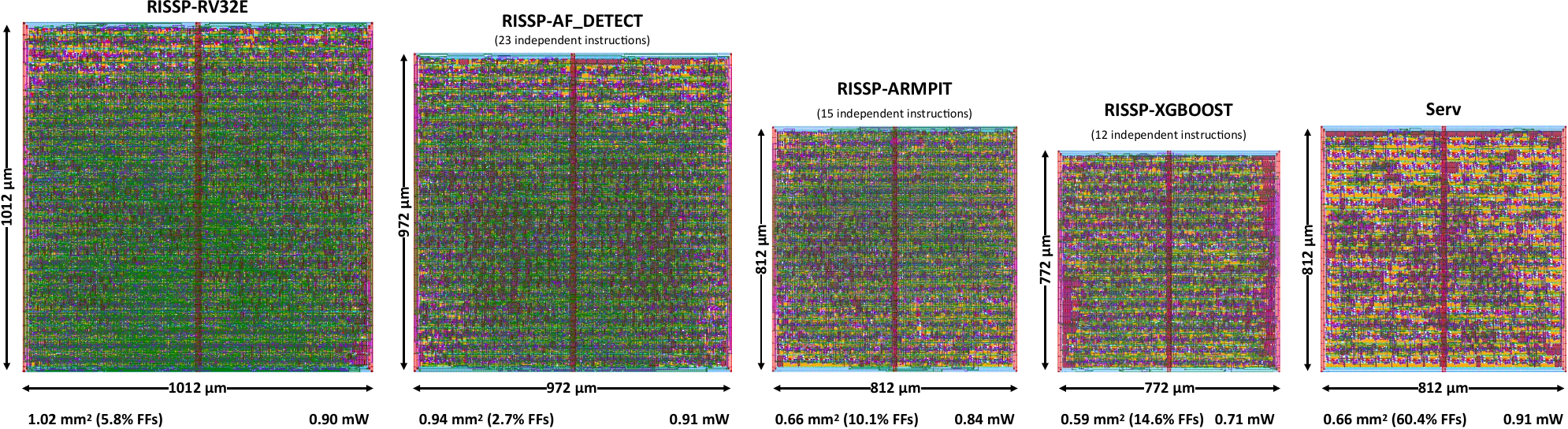}
    \caption{Layouts of the five implemented processors are shown. The leftmost layout corresponds to \textit{RISSP-RV32E}. Next are the three RISSPs for \textit{af\_detect}, \textit{armpit}, and \textit{xgboost} — each representing an extreme-edge application. The rightmost layout is \textit{Serv}. For each layout, the X and Y dimensions of the die area are shown in \textmu m, along with the die area (in mm²) and percentage of FFs being used in each design, on the bottom-left of each layout. The total power consumption (in mW) is indicated on the bottom-right. Additionally, for the three RISSPs, the number of distinct instructions implemented in each design is annotated.}
    \label{graph:impl}
\end{figure*}

The physical implementation is initially started at the maximum clock frequency of each processor found at the synthesis stage. Due to clock tree insertion and routing delays, the implementation cannot achieve the maximum clock frequency for all processors. We follow an iterative implementation process to incrementally reduce the clock frequencies until all processors became functional at 300 kHz. All processors are implemented at this frequency using typical and nominal corner conditions with a supply voltage of 3V.

\textbf{Figure \ref{graph:impl}} shows the layouts of the three RISSPs, \textit{RISSP-RV32E} and \textit{Serv}. \textit{RISSP-af\_detect} achieves an ~8\% reduction in area compared to \textit{RISSP-RV32E} whilst \textit{RISSP-armpit}  has the same area as \textit{Serv} but reduces the area by \textasciitilde35\% with respect to \textit{RISSP-RV32E}. \textit{RISSP-xgboost} achieves even further reduction by \textasciitilde11\% and  \textasciitilde42\% over \textit{Serv} and \textit{RISSP-RV32E}. This is because \textit{Serv} has a high proportion of FFs (i.e., 60\%), and therefore requires more clock tree buffers, which increases its area despite the fact that the synthesis results indicating that \textit{Serv} is smaller than \textit{RISSP-xgboost}.

The two baselines \textit{RISSP-RV32E} and  \textit{Serv} have almost the same power consumption in spite of \textit{Serv} being \textasciitilde35\% smaller than \textit{RISSP-RV32E}. This is again because of the high proportion of FFs in \textit{Serv} whereas FFs take up only \textasciitilde6\% of \textit{RISSP-RV32E}. \textit{RISSP-af\_detect} consume almost the same power as \textit{Serv} and \textit{RISSP-RV32E} whilst \textit{RISSP-armpit} and \textit{RISSP-xgboost} consume \textasciitilde8\% and \textasciitilde21\% less power than \textit{Serv} and \textit{RISSP-RV32E}, respectively.

\begin{figure*}[!ht]
    \centering
    \includegraphics[width =\textwidth]{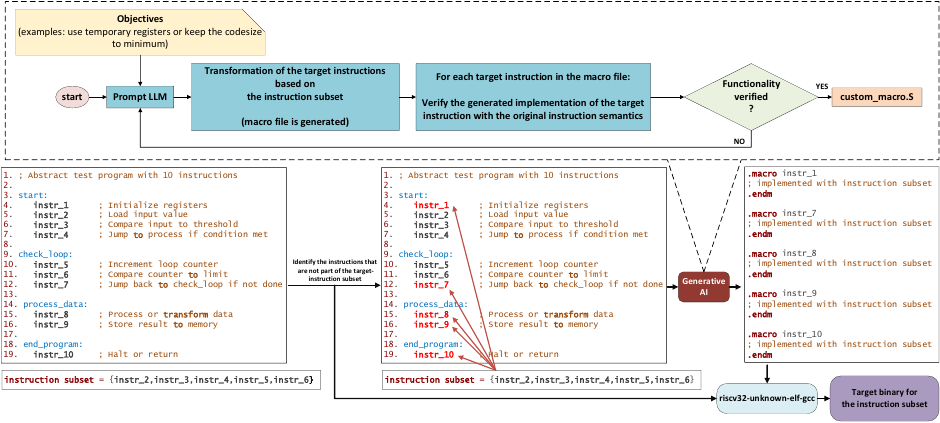}
    \caption{Generative AI based tool to implement code retargeting for RISSPs used in long-lasting extreme edge applications}
    \label{fig:compilation_flow}
\end{figure*}

%% file: src/compiler.tex
\section{Generative AI based Code Retargeting}
\label{sec:recompilation}

Although the concept behind the RISSP generation is to rely on the RISC-V software ecosystem rather than generating a custom software toolchain, long-lasting extreme edge applications need frequent software updates because the target applications may change. For instance, applications need to be recompiled with different compiler optimization flags, which may generate new instructions that are not supported by the RISSP under consideration.

There are two possible approaches to address this scenario. The first approach focuses on modifications at the compiler backend that needs to be modified to receive the instruction subset implemented in the RISSP and constrain the assembly code generation to this subset only. This method is not trivial and requires significant changes in the instruction selection and code generation phases of the compiler backend. The second approach relies on the assumption that the assembly code is generated using standard compiler toolflows and a separate tool is responsible to transform the generated assembly into a version that uses the instruction subset supported by the RISSP.

We follow the second approach where we develop a tool that uses generative AI to transform the assembly code generated by the compiler targeting the full RISC-V ISA (e.g., \texttt{riscv32-unknown-elf-gcc}) to assembly code that uses \textit{only} the instructions in the instruction subset. This is an important step towards reusability of RISSPs after fabrication. \textbf{Figure \ref{fig:compilation_flow}} presents the retargeting framework along with a simple example. Given an application and an instruction subset, the tool is able to identify the instructions that are not included in the instruction subset. Then, with the use of ChatGPT RISC-V Assembly plugin \cite{chatgpt_assembly}, every instruction that is not part of the instruction subset is transformed into a set of instructions that are included in the instruction subset and is defined as an independent \texttt{macro}. The \texttt{macros} of the transformed instructions are stored in a \texttt{macro.S} file. Finally, the application is recompiled using the standard \texttt{riscv32-unknown-elf-gcc} with the \texttt{macro.S} file.

Initially, we provide the instruction subset and the instructions that need to be transformed to the ChatGPT RISC-V Assembly plugin. Objectives that restrict the solution space can be added based on the user preferences. For example, such objectives could be ``\textit{allow the use of temporary registers to provide a solution}'', or, ``\textit{keep the codesize to minimum as you transform the instructions}''. After that, the process of generating \texttt{macros} for the non-supported instructions from the instruction subset begins. Then, the \texttt{macros} are verified against the original semantics of the transformed instruction with custom test cases. If the LLM generates a \texttt{macro} that cannot be functionally verified, the macro is rejected, and then another \texttt{macro} is requested from the LLM. This continues until a valid \texttt{macro} is generated. In our experiments, a valid \texttt{macro} can be generated on less than 10 attempts.

\begin{figure}[!h]
    \centering
    \includegraphics[width =0.6\columnwidth]{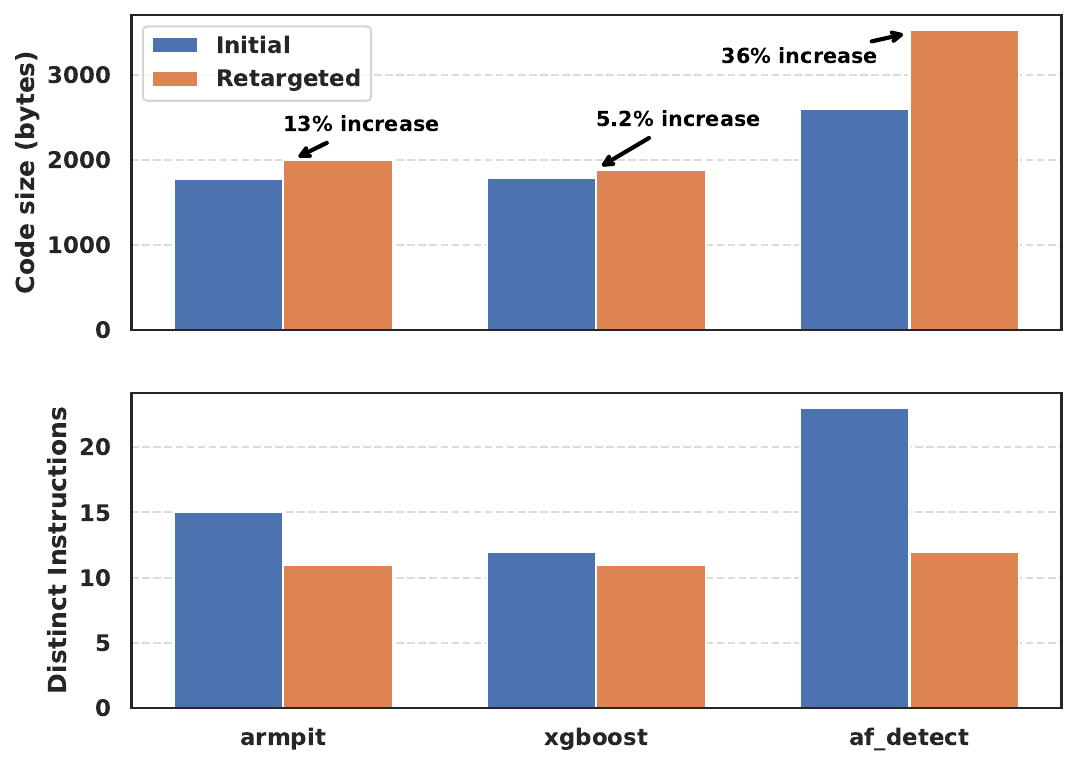}
    \caption{Comparison of code size (bytes) and the number of distinct instructions between the initial binary and the retargeted binary using LLM}
    \label{fig:llm_analysis}
\end{figure}

To evaluate the above framework, we define a minimal instruction subset of twelve instructions \texttt{\{addi, add, and, xori, sll, sra, jal, jalr, blt, bltu, lw, sw\}} from which other instructions can be reproduced. \textit{Armpit}, \textit{xgboost} and \textit{af\_detect} can be defined as ``long-lasting extreme edge applications'', as discussed in \textit{Section \ref{sec:evaluation}}, and for this experiment they are chosen as a case study to show the potential of the framework. The framework receives as input the initial version of the target applications compiled with \texttt{-O2}, and attempts to transform the non-supported instructions of the target applications to functionally relevant \texttt{macro} based on the instruction subset.

\textbf{ Figure \ref{fig:llm_analysis}} shows that the code size after the recompilation of the applications are 13\%, 5.2\% and 36\% larger than the original version for \textit{armpit}, \textit{xgboost} and \textit{af\_detect}, respectively. The number of distinct instructions shows minimal variation for \textit{armpit} and \textit{xgboost}, while the number of distinct instructions for \textit{af\_detect} is reduced from 23 to 12. This experiment demonstrates the feasibility of this approach to efficiently retarget code but with an overhead of larger code size and potentially slower runtime of the application.

%% file: src/conclusions.tex
\section{Conclusions}
\label{sec:conclusions_and_fw}

We have proposed a methodology for automatically generating a single-cycle 32-bit RISC-V instruction subset processor (RISSP) to be deployed as a FlexIC to meet the computing requirements of the extreme edge. The methodology constructs a RISSP supporting only the instruction subset identified during the characterization of an extreme edge application or domain. The semantics of each instruction in the RISC-V ISA are translated into a formally verified and discrete, hardware block and stored in a library. The hardware blocks of the instruction subset are extracted from the library, stitched together to build the core of the RISSP, which is then connected to the rest of the pre-verified processor components to generate the RISSP. The RISSP  methodology reduces the time-to-market of custom processors by integrating verification as a primary design principle. The synthesis results have shown that RISSPs can achieve 8-to-43\% reduction in area and 3-to-30\% reduction in power compared to a processor supporting the full RISC-V ISA, and are on average about 40 times more energy efficient than \textit{Serv}- the world’s smallest 32-bit RISC-V processor. When implemented as FlexICs, the RISSPs targeting extreme edge applications can achieve up to 42\% and 21\% less area and power.

The proposed methodology can support custom instructions that can be added to the pre-verified ISA hardware library like other instructions in the ISA. Also, the methodology can be extended to generate pipelined RISSPs if higher clock frequencies are required but is not suitable for generating complex processors that have deeper superscalar and out-of-order execution pipelines. Extreme edge applications do not require such high performance processors, so our focus has been to develop a microcontroller-class RISSP that is a single-cycle scalar in-order execution machine providing adequate performance for applications at extreme edge.